\documentclass[floatfix,epsf,prb,twocolumn,amsmath,amssymb,showpacs,
superscriptaddress]{revtex4}
\usepackage[english]{babel}
\usepackage{graphics}
\usepackage{amsmath}
\usepackage{dsfont}
\usepackage{graphicx}
\usepackage[figtopcap]{subfigure}
\usepackage{enumerate}
\begin{document}\setlength{\unitlength}{1mm}
\newcommand{\ud}{\mathrm{d}}
\newcommand{\kvec}[2]{\begin{pmatrix} #1 \\ #2 \end{pmatrix} }
\newcommand{\rvec}[2]{\begin{pmatrix}#1 & #2 \end{pmatrix} }
\newcommand{\matt}[4]{\begin{pmatrix} #1 & #2 \\ #3 & #4 \end{pmatrix} }
\newcommand{\pdje}[1]{\partial_{#1}}
\newcommand{\ve}{\varepsilon}
\title{Extra Dirac points in the energy spectrum for superlattices on single-layer graphene}

\author{M. Barbier}
\affiliation{Department of Physics, University of Antwerp, Groenenborgerlaan 171, B-2020 Antwerpen, Belgium\\}
\author{P. Vasilopoulos}
\affiliation{Department of Physics, Concordia University, 7141 Sherbrooke Ouest, Montr\'eal, Quebec, Canada H4B 1R6\\}
\author{F. M. Peeters}
\affiliation{Department of Physics, University of Antwerp, Groenenborgerlaan 171, B-2020 Antwerpen, Belgium\\}
\begin{abstract}
We investigate the emergence of extra Dirac points in the electronic 
structure of a periodically spaced barrier system, i.e., a superlattice, on 
single-layer graphene, using a Dirac-type Hamiltonian. Using square barriers 
allows us to find  analytic expressions for the occurrence and location of 
these new Dirac points in {\bf k} space and for the renormalization of the 
electron velocity near them in the low-energy range. In the general case of 
unequal barrier and well widths the new Dirac points move away from the 
Fermi level and for given heights of the potential barriers there is a minimum 
and maximum barrier width outside of which the new Dirac points disappear. The 
effect of these extra Dirac points on the density of states and on the 
conductivity is investigated.
\end{abstract}
\maketitle
\section{Introduction}\label{sec1}
Graphene, a one-atom thick layer of carbon atoms, has been a topic of intense 
study since it's experimental realization\cite{geim1} in 2004. Interest  in 
graphene results, in particular, from 
the prediction that  carriers in it  behave as massless, chiral Dirac fermions, 
moving in a two-dimensional (2D) plane and  described by a Dirac-type 
Hamiltonian. This model predicts 
unusual electronic properties such as the gapless electronic spectrum, the 
perfect transmission at normal incidence through any potential barrier, i.e., 
the Klein paradox \cite{klein,kats}  which was recently addressed 
experimentally\cite{kex}, the zitterbewegung  recently verified \cite{zit}, 
etc., see Ref.~\onlinecite{cast} for a recent review.

Motivated by all these properties, condensed matter physicists started extending 
the known properties of a two-dimensional electron gas (2DEG) in semiconductor 
materials to the relativistic 2D fermions (Dirac electrons) in graphene. Another 
particularly interesting system to consider is the application of a periodic 
potential  to graphene, that is, a superlattice (SL), which under special 
conditions leads to collimation of electron beams\cite{parkanisotropy,parksgs,nori}.

As found recently in Ref. \onlinecite{ho},  using a tight-binding formalism, 
the dispersion relation for such a SL can reveal extra 
Dirac points at the Fermi level \cite{hosummary}. Close to the Fermi level the 
electronic properties of graphene are well described by the massless 
2D Dirac equation. 
In two recent studies, Ref. \onlinecite{parkwiggles} and 
\onlinecite{breywiggles}, an exact condition was found for the emergence of 
extra Dirac points (zero modes)  in the presence  of  a sinusoidal or 
square-wave SL potential. However, both studies are not able to describe the 
character and spatial distribution in ${\bf k}$ space of these new Dirac points, 
as they expand the spectrum for small $k_y$. In this work we describe  under 
which condition this is possible and  also where  these extra Dirac points arise 
in the electronic structure of massless Dirac fermions in single-layer graphene 
when  a square-wave periodic potential is 
applied. Further, we analytically investigate the anisotropic renormalization of 
the group velocities at these new Dirac points, 
and find that the degree of the renormalization depends on the 
parameter $u \propto V_0 L$, which is linear in the barrier height $V_0$ and 
period $L$ of the SL, in the sense that, $v_x > v_y$ holds for $u$ such that an 
extra Dirac point arises while for very high values $u$ we have $v_x < v_y$. 
Moreover, we  also consider the case of unequal barrier and well widths, 
not treated previously,  that results in a qualitatively different 
electronic spectrum. 

We organize the paper as follows. In Sec.~\ref{sec2} we introduce our model. In 
Sec.~\ref{sec3} we investigate the emergence of the extra Dirac points, 
approximate the implicit 
dispersion relation for small energies, and take a closer look at the group 
velocity near the extra Dirac points. Further we investigate the influence of 
the features of the spectrum on  the density of states and conductivity. 
Finally, we make a summary and  concluding remarks in Sec.~\ref{sec4}.
\section{Model}\label{sec2}
\begin{figure}[ht]
  \begin{center}
	\includegraphics[height=3cm]{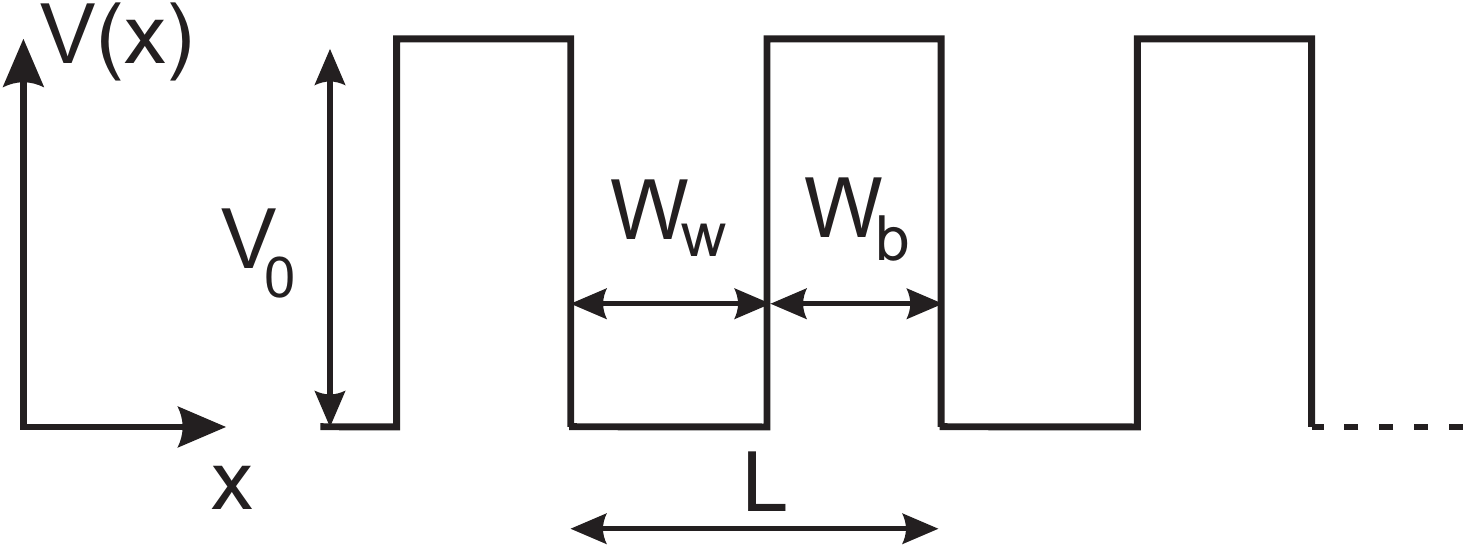}
    \end{center}
	\vspace{-0.7cm}
    \caption{Schematics of the superlattice potential V(x) consisting of square barriers.}\label{fig2_1}
\end{figure}
We describe the electronic structure of an infinitely large flat graphene flake 
by the nearest-neighbour, tight-binding model and consider solutions with energy 
and wave vector close to the K point. 
The relevant Hamiltonian is 
$\mathcal{H} = v_F \vec{\sigma}\cdot \hat{\vec{p}} + \mathds{1} V$, with 
$\mathds{1}$ the $2\times 2$ unit matrix. Explicitly $\mathcal{H} $ 
is given by
\begin{equation}\label{eq2_1}
	\mathcal{H} = \matt{V}{-i v_F \hbar (\pdje{x}-i \pdje{y})}{-i v_F \hbar (\pdje{x}+i \pdje{y})}{V},
\end{equation}
where $\vec{p}$ is the momentum operator 
and $v_F \approx 10^6 m/s$ the Fermi velocity. In the presence of 
a one-dimensional (1D) square-wave potential $V(x)$, such  as the one shown in 
Fig. 1, the equation $(\mathcal{H} - E)\psi = 0$ admits solutions of the form 
$\psi(x) e^{i k_y y}$ with
\begin{equation}\label{eq2_2}
	\psi(x) = \kvec{1}{s e^{i \phi}}e^{i \lambda x}, \quad  \psi(x) = \kvec{1}{-s e^{-i \phi}}e^{-i \lambda x},
\end{equation}
$\lambda = [(\ve-u(x))^2 - k_y^2]^{1/2}$,\,\,$\tan \phi = k_y/\lambda$, 
$s = sign(\ve-u(x))$,  $\ve = E/v_F \hbar$, and $u(x) = V(x)/v_F \hbar$; the 
parameters $\ve $ and $u(x)$ are in units of inverse length.

\subsection{Square-barrier superlattice}\label{sec2A}
We consider an infinite number of periodically spaced barriers, 
as shown in Fig. \ref{fig2_1}, with unit cell length $L$ and barrier (well) 
width $W_b$ ($W_w$). It is convenient to introduce 
the dimensionless variables $\ve \rightarrow \ve L$, $k_y \rightarrow k_y L$, 
$k_x \rightarrow k_x L$, $u \rightarrow u_0 L = V_0 L/v_F \hbar$, 
$x \rightarrow x / L$, $W_b \rightarrow W_b / L$, and 
$W_w \rightarrow 1-W_b / L$. The wave function of this periodic system is a 
Bloch function and the transfer matrix $\mathcal{T}$ pertinent to it 
leads to an expression for the dispersion relation, see Appendix \ref{app1}. 
For $|k_y|<|\ve_w|$  and $\ve_w$ as in Eq. (\ref{eq2a_4}), the transfer matrix 
$\mathcal{T}$ can be written as \cite{griffiths}
\begin{equation}\label{eq2a_1}
	\mathcal{T} = \matt{w}{z}{z^*}{w^*}; 
\end{equation}
then the dispersion relation becomes
\begin{equation}\label{eq2a_2}
	\cos (k_x) = \Re\{e^{-i \lambda} w\},
\end{equation}
with $w$ given by
\begin{equation}\label{eq2a_3}
	w = e^{i \lambda W_b} \left[ \cos \Lambda W_b - i G \sin \Lambda W_b \right]
\end{equation}
and  
\begin{eqnarray}\label{eq2a_4}
\nonumber
	&& \ve_w = \ve + u W_b, \,\, \ve_b = \ve - u W_w,\,\, G = (\ve_w \ve_b - k_y^2)/\lambda \Lambda,\\*
	&& \lambda = [\ve_w^2 - k_y^2]^{1/2}, \quad \Lambda = [\ve_b^2 - k_y^2]^{1/2}. 
\end{eqnarray} 
Writing the rhs of Eq. (\ref{eq2a_2})   explicitly gives  
\begin{equation}\label{eq2a_5}
	\cos k_x = \cos\lambda W_w \cos\Lambda W_b - G \sin\lambda W_w \sin\Lambda W_b.
\end{equation}

Although this derivation is only correct for $|k_y|<|\ve_w|$, Eq. (\ref{eq2a_5}) 
is also valid beyond this limitation, see Ref. \onlinecite{barb}. 
From Eq. (\ref{eq2a_5}) it can be seen that the dispersion relation possesses 
the symmetry property $\ve \rightarrow - \ve$ for $W_b \leftrightarrow W_w$. 
The asymmetric spectrum is not unexpected because the symmetry of the potential 
about the Fermi level is lost  for $W_b \neq 1/2$.  For $W_b = 1/2$ we have
\begin{equation}\label{eq2a_6}
	\cos k_x = \cos \frac{\lambda}{2} \cos \frac{\Lambda}{2} - G \sin \frac{\lambda}{2} \sin \frac{\Lambda}{2},
\end{equation}
where $\ve_w = \ve + u/2$ and $\ve_b = \ve - u/2$. For this interesting case, 
the potential  possesses particle-hole symmetry   and the extra Dirac points 
originate at the Fermi level; we will show their  arrangement, in 
${\bf k}$ space, in Sec. \ref{sec3A}.
\begin{figure}[ht]
	\begin{center}
		\subfigure{\includegraphics[height=4.5cm]{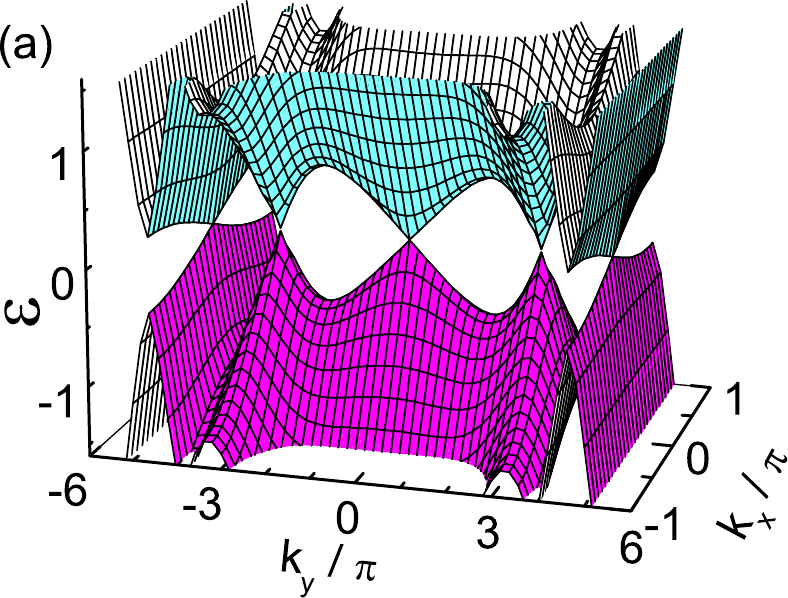}} 
		\subfigure{\includegraphics[height=4.5cm]{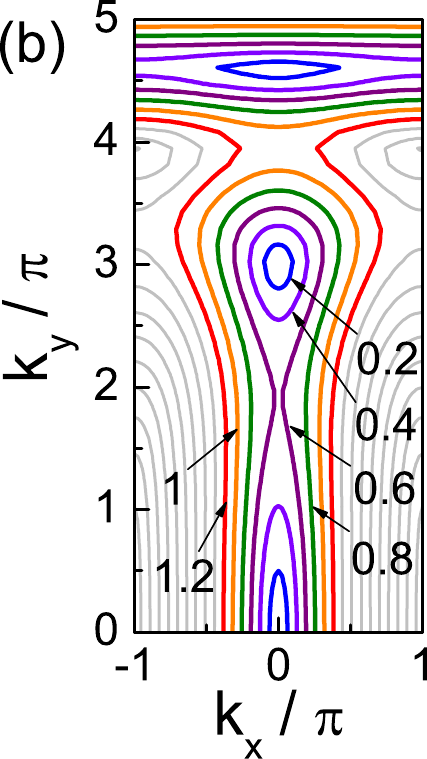}} 
	\end{center}
	\vspace{-0.7cm}
	\caption{(Color online) (a) Valence and conduction bands 
	of the spectrum of a SL with square 
	barriers of  width $W_b = 1/2$ and height $u = 10\pi$. 
	(b) Contourplot of the conduction band.}
	\label{fig2}
\end{figure}

In Fig. \ref{fig2}(a) the spectrum resulting from Eq. 
(\ref{eq2a_6}) for equal barrier and well widths, i.e., for $W_w= W_b = 1/2$, is 
plotted for $u = 10 \pi$. 
As can be seen, the spectrum is symmetric 
about the Fermi level; there are two extra Dirac points on both sides of the 
main Dirac point, and their velocities are renormalized. 
The anisotropic behaviour of the new Dirac cones is clearer  
in the projection of the conduction band shown in Fig. \ref{fig2}(b). 
Further details  about the  renormalization of the velocities will be given in 
Sec. \ref{sec3C}.
\begin{figure}[ht]
\hspace{-0.02\linewidth}
\begin{minipage}{0.50\linewidth}
\centering
\subfigure{\includegraphics[height=4cm,width=5.3cm]{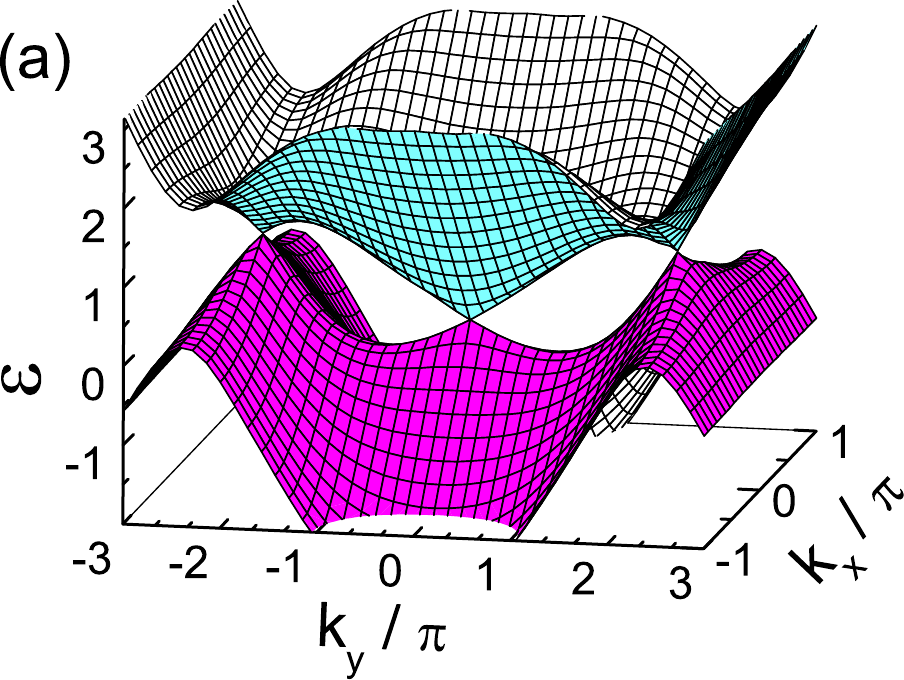}} 
\end{minipage}
\hspace{0.08\linewidth}
\begin{minipage}{0.40\linewidth}
\centering
\subfigure{\includegraphics[height=2.3cm]{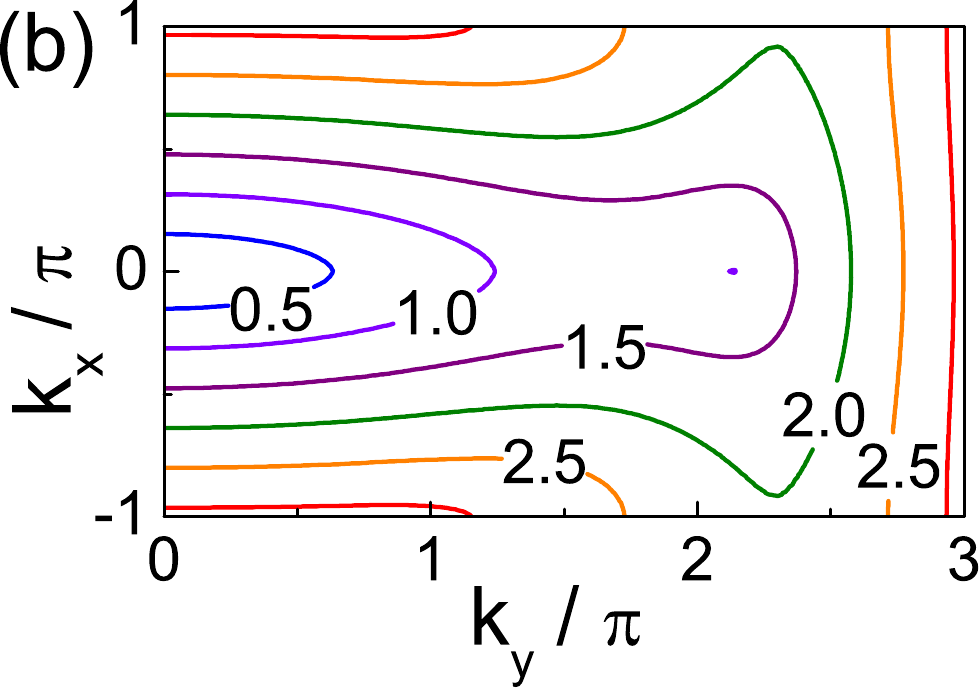}}\\ 
\subfigure{\includegraphics[height=2.3cm]{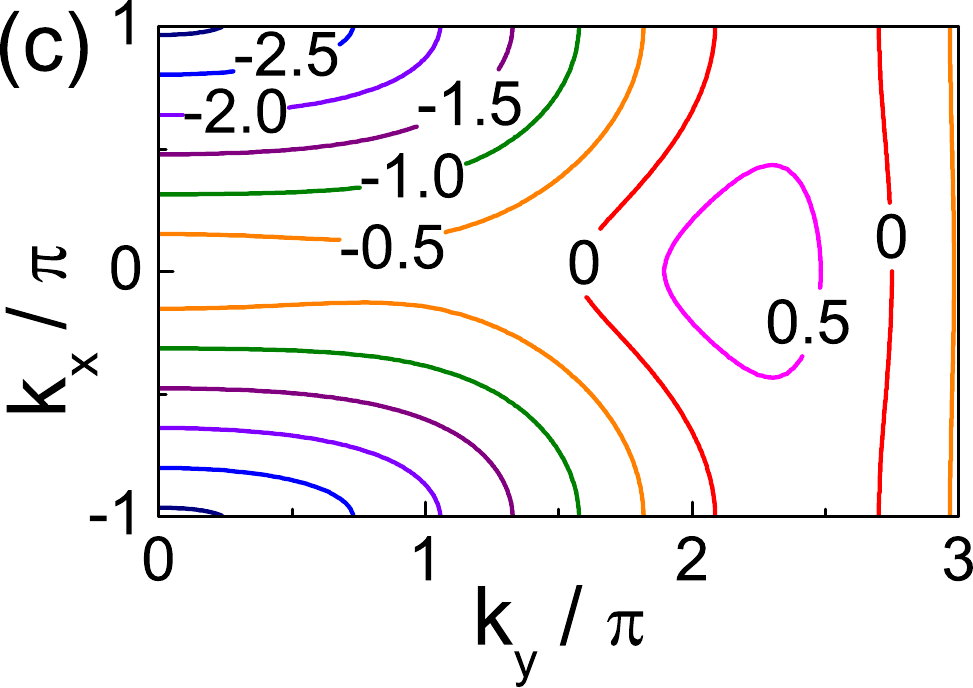}} 
\end{minipage}
	\caption{(Color online) 
	(a) Valence and conduction bands of the spectrum of 
	a SL with barriers of  width $W_b = 0.4$ and height $u = 6\pi$. (b) and (c): 
	projection contours of the conduction and valence band, respectively, on 
	the $(k_x, k_y)$ plane.}\label{fig10}
\end{figure}

For unequal barrier and well widths the spectrum is shown in 
Fig. \ref{fig10}(a) for $W_b = 1-W_w = 0.4$ and  
$u = 6 \pi$. The spectrum is no longer symmetric about the Fermi level, the two 
extra Dirac points are shifted in energy relative to the main  point  and their 
velocities are renormalized. The location of the extra Dirac points will be 
investigated in Sec. \ref{sec3}.
\section{Electronic structure}\label{sec3}
\subsection{Appearance of extra Dirac points}\label{sec3A}
In order to find the location of the Dirac points 
we assume $k_x = 0$, $\ve = 0$, and $W_b = W_w = 1/2$ in Eq. (\ref{eq2a_6}). 
Then Eq. (8) becomes
\begin{equation}\label{eq3_2}
	1 = \cos^2 \lambda/2 + \big[(u^2/4 + k_y^2)/(u^2/4 - k_y^2)\big] \,\sin^2 \lambda/2,
\end{equation} 
which has solutions for  $u^2/4 - k_y^2 = u^2/4 + k_y^2$  or 
$\quad \sin^2 \lambda/2 = 0$. 
For the first possibility $k_y = 0$ is the only solution and corresponds to the 
usual Dirac point. The second possibility leads 
to $\lambda/2 = j \pi$  with  $j \neq 0$, because $ \lambda= 0$ makes the 
denominator $u^2/4 - k_y^2 = \lambda^2$ vanish 
and does not lead to a solution. For 
$\lambda/2 = j \pi$  we have 
\begin{equation}\label{eq3_3} 
\hspace*{-0.27cm}k_{y,j\pm} = \pm\sqrt{
	\frac{u^2}{ 4} - 4 j^2 \pi^2}  
	= \pm\sqrt{ 
	\Big(\frac{V_0}{2 \hbar v_F}\Big)^2 
	- 	\Big(\frac{2 j \pi}{L}\Big)^2},  
\end{equation}
where we reinserted the  dimensions after the second  equality  sign.  As such,  
Eq. (\ref{eq3_3}) describes the spatial arrangement  
of the extra Dirac points along the $k_y$ axis. Also, it clearly shows how many 
points we have at particular values of $u$, 
namely $2 \times (u \mod 4 \pi)$, 
and where they are  located in 
{\bf k} space.  Each time $u$ becomes  a multiple of $4 \pi$ a new pair of Dirac 
points is generated for $k_y = 0$.  The condition 
$j \neq 0$ gives us a threshold value of 
$u = 4 \pi$ for the emergence of the first pair. 
The integer $j$ denotes the $j$th  
extra Dirac point, so the outer extra Dirac points have $j = 1$ as they are 
generated first. 
\begin{figure}[ht]
	\begin{center}
	  	\subfigure{\includegraphics[height=3.5cm]{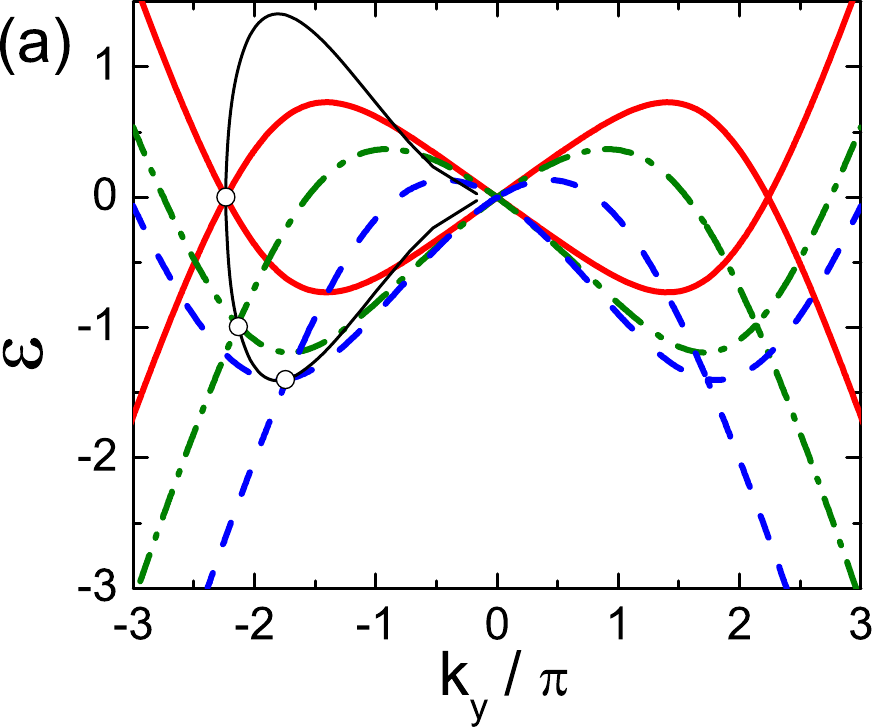}} 
	  	\subfigure{\includegraphics[height=3.5cm]{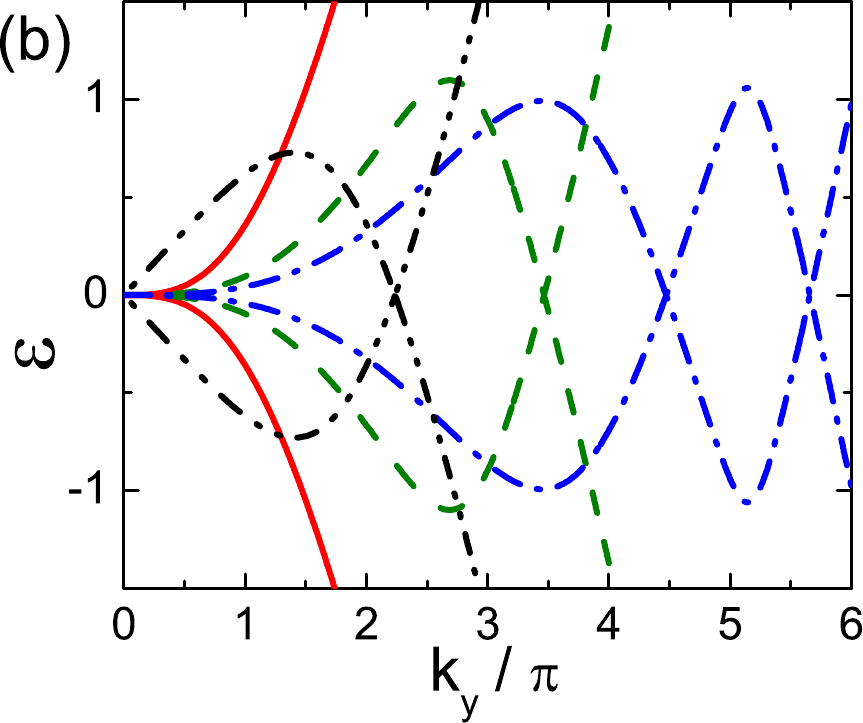}} 
	\end{center}
	\vspace{-0.7cm}
	\caption{(Color online) Slices of the SL spectrum along $k_y$ with $k_x = 0$ 
	and $u = 6\pi$. (a) The solid red, dash-dotted green,  and dashed blue  
	curves correspond to barrier widths $W_b = 0.5, 0.6$, and $0.7$ 
	respectively. The thin black line is the curve on which the extra Dirac 
	point, to the left of the main one at $k_y=0$, is located for various $W_b$. 
	Only the new points to the left of the main one are shown. 
		(b) As in (a) for fixed $W_b = 0.5$. 
	The solid red, dot-dot-dashed black, dashed green, and dash-dotted blue  
	curves are for different values of the barrier height such that 
	$u/2 = 2\pi, 3\pi, 4\pi$, and $6\pi$, respectively.
 }\label{fig3}
\end{figure}

In Fig. \ref{fig3}(a) we show slices of the SL spectrum along $k_y$ for 
$k_x = 0$ and $u = 6\pi$. 
The solid red, dash-dotted green,  and dashed blue dashed curves correspond to 
barrier widths $W_b = 0.5, 0.6$, and $0.7$ respectively.  The thin black line is 
the curve on which the extra Dirac points, 
on the left of the main one at $k_y=0$,  are located for various $W_b$. 
In Fig. \ref{fig3}(b) we show slices of the spectrum along $k_y$ for $k_x = 0$. 
The solid red, dot-dot-dashed black, dashed green, and dash-dotted blue  
curves are for different values of the barrier height 
such that $u/2 = 2\pi, 3\pi, 4\pi$, and $6\pi$ respectively. For  values of 
$u/2$ which are multiples of $2 \pi$, new Dirac points are generated. 
Interestingly, if new extra points are to arise, the dispersion becomes almost 
flat along the $k_y$ axis at the Dirac point, i.e,  collimation occurs. We will 
come back to this issue in Sec. \ref{sec3C}.

\emph{Unequal well and barrier widths}.  
We return  to the more general case of unequal well and barrier widths for which 
$W_b \neq 1/2$.  It is more difficult to   locate the extra Dirac points which  
no longer occur at the Fermi level as seen from 
the green and blue curves in  Fig. \ref{fig3}(a) showing slices of spectra from 
Eq. (\ref{eq2a_5}) for $k_x = 0$. 
By means of the symmetry  $\ve \rightarrow - \ve$ for $W_b \leftrightarrow W_w$, 
we know the complementary plots for $W_b \rightarrow 1 - W_b$. 
As can be seen, the extra Dirac points shift mainly down (up)  in energy as  
$W_b$ increases (decreases). 
To find their coordinates $(\ve,k_x = 0, k_y)$ we 
assume $\sin(\lambda W_w)$=$\sin(\Lambda W_b)=0$ and 
$\cos(\lambda W_w) = \cos(\Lambda W_b) = \pm 1$\cite{noimplicit2}. 
This gives (Appendix \ref{app2})
\begin{equation}
\begin{aligned}
	& \ve_{j,m} = \frac{u}{2}(1-2 W_b) + \frac{\pi^2}{2 u}\left( \frac{j^2}{W_w^2}-\frac{(j+2 m)^2}{W_b^2}\right),\\
	& {k_y}_{j,m} = \pm  
	\Big[(\ve_{j,m} + u W_b)^2 -(j \pi / W_w)^2\Big]^{1/2},
\end{aligned}
\end{equation}
where $j$ and $m$ are integers. This method also shows higher and lower crossing 
points if $m \neq 0$. In Fig. \ref{fig3}(a) the extra Dirac points on the left, 
obtained with this method, are 
indicated by open circles and the thin black curve shows their trajectory 
in $(E,k_y)$ space as the width $W_b$ varies. For a particular $u$ there is a 
minimal width $W_b$ (and a corresponding maximal width 
$W_b \rightarrow 1 - W_b$) below (above) 
which  the various extra Dirac points disappear. 
In Fig. \ref{fig3}(a) the ``$ $Dirac cones'' at these 
crossing points for $m = 0$ are not only reshaped with a renormalized 
anisotropic velocity  but, as 
shown  by the blue dashed curve,  the ``$ $extra Dirac point'' is not at a local 
minimum (maximum) of the conduction (valence) band. 

\subsection{Analytical expression for the spectrum for small energies $\ve$}\label{sec3B}
As the purpose is to have a closer look at the behaviour of the extra Dirac 
points and we cannot prohibit $k_y$  from being large, 
we expand Eq. (\ref{eq2a_6}) for small energies, up to second order in  $\ve$, 
and obtain the following explicit dispersion relation 
\begin{equation}\label{eq6_1}
\hspace*{-0.37cm}	\ve_\pm = \pm 
	\left[{\frac {4|a^2|^2\,\big[k_y^2 \sin^2 (a/2) + a^2 \sin^2(k_x/2)\big]}{k_y^4 a \sin a + a^2 u^4/16 - 2 k_y^2 u^2 \sin^2 (a/2) 
	}}\right]^{1/2},
\end{equation}
with $a = [u^2/4 - k_y^2]^{1/2}$.  

If we only need the behaviour of the spectrum near the K point (for small $k_x$ 
and $k_y$), it suffices to make an expansion for small $\ve$ and $k_y$ in 
Eq. (8), up to third order in products of $\ve$ and $k_y$ since this is the 
first order with an energy dependence. The result is
\begin{equation}
	2 \cos k_x - 2 + \ve^2 - k_y^2 \,\sin^2(u/4)/(u/4)^2 = 0.
\end{equation}
Then we solve for the energy $\ve$ and  obtain
\begin{equation}\label{eq5_2}
	\ve \approx \pm  
	\big[4\sin^2 k_x/2 + k_y^2\, \sin^2(u/4)/(u/4)^2\big]^{1/2}.
\end{equation}

\begin{figure}[ht]
  \begin{center}
	\subfigure{\includegraphics[height=3.6cm]{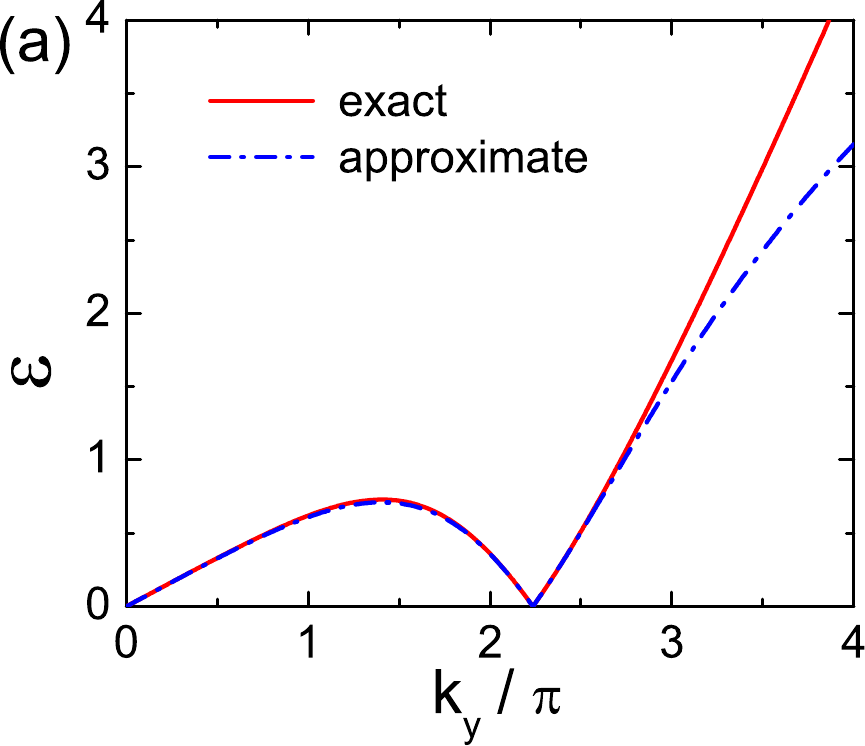}} 
	\subfigure{\includegraphics[height=3.5cm]{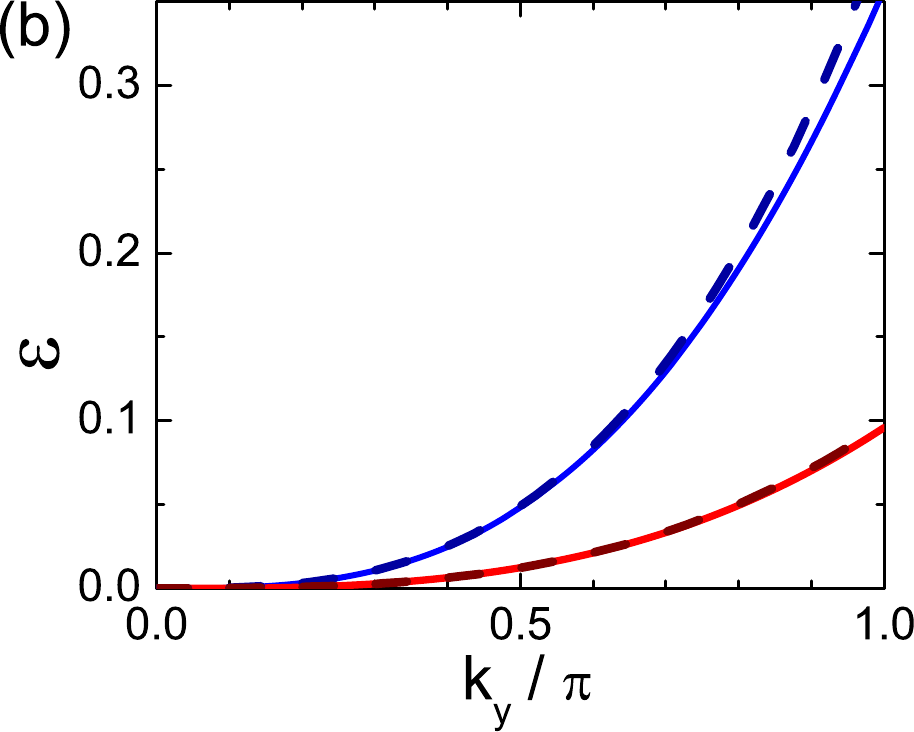}} 
  \end{center}
  \vspace{-0.7cm}
  \caption{(Color online) (a) Plot of the dispersion relation resulting from 
  Eq. (\ref{eq6_1}) (dashed  blue curve) versus the exact one  obtained 
  from Eq. (\ref{eq2a_6}) (solid red curve) for $k_x = 0$ and $u = 6 \pi$. 
  (b) The 
   solid blue and red curves show, respectively, the zoomed-in plots of the 
   solid red and dashed green curves of Fig. \ref{fig3}(b), i.e., for $u=4\pi$ 
   and $u=8\pi$, respectively, and $W_b = W_w =1/2$. The approximation of these 
   curves by Eq. 
   (\ref{eq_approxfig6}) are the dashed curves.}\label{fig6}
\end{figure}

In Fig. \ref{fig6}(a) we show $\ve$ from  Eq. (\ref{eq6_1})  
and compare it with the exact 
dispersion relation, for $k_x = 0$, following from Eq. (\ref{eq2a_6}). The 
expansion (\ref{eq6_1}) is rather good for low energies near the extra Dirac 
points; accordingly, we will use  Eq. (\ref{eq6_1}) to further assess their 
behaviour.

\subsection{Anisotropic velocity renormalization\\ at the (extra) Dirac point(s).}\label{sec3C}

The spectrum in the low-energy range consists of two kinds of valleys, 
one near 
the main Dirac point and the other near the extra Dirac points. Near the 
original Dirac point the spectrum is almost linear, 
perpendicular to the barriers, and zero parallel to them, 
whereas near the extra Dirac points the 
situation can be reversed 
depending on the height of the barriers.

\begin{figure}[ht]
  \begin{center}
  	\subfigure{\includegraphics[height=5cm]{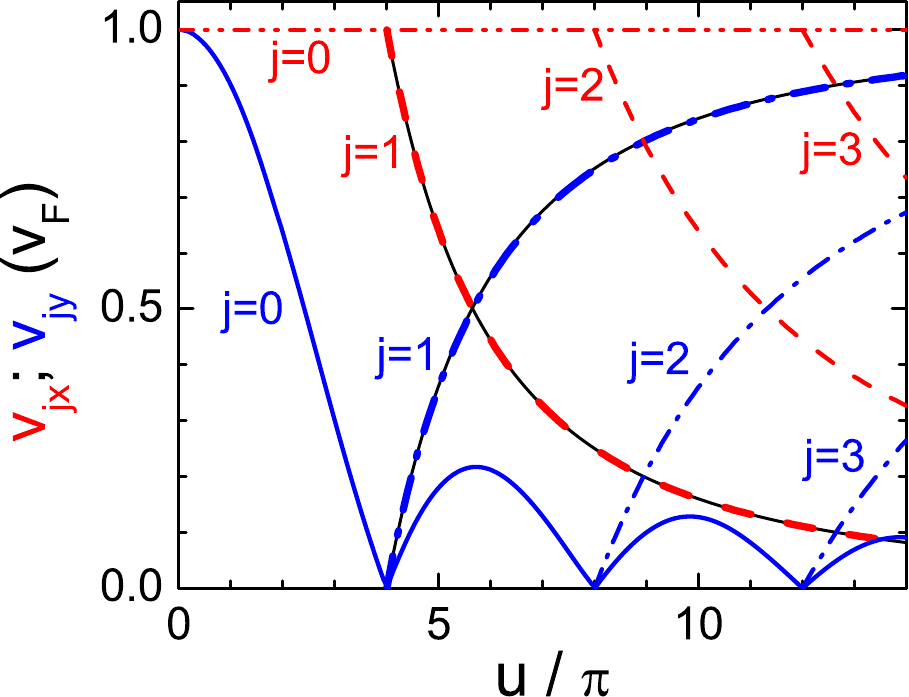}}
  \end{center}
	\vspace{-0.7cm}
	\caption{(Color online) 
  Velocities $v_{0x}$ and $v_{0y}$ (dash-dot-dotted red and solid blue curves, 
  respectively), vs $u$ at the original Dirac point and $v_{jx}$ and $v_{jy}$ 
  (dashed red and dash-dotted blue curves, respectively), given by Eqs 
  (\ref{eq_vx}) and (\ref{eq_vy}), vs $u$ at the extra Dirac points 
  $j = 1, 2, 3$. The thin black curves for the  $j = 1$ Dirac point, are 
  obtained numerically from Eq. (\ref{eq2a_6}).
  }\label{fig6f}
\end{figure}

\emph{Group velocity along the $x$-axis at the Dirac point.}  
To compare with the collimation found by Park {\it et al.}\cite{parksgs}, 
we notice that in Fig. \ref{fig3}(b), for the 
solid red, dashed green, and dash-dotted blue curves, corresponding to barrier 
heights which are multiples of $4 \pi$, the dispersion 
becomes more 
flat for small $k_y$. From Eq. (\ref{eq5_2}) we could already expect that, to 
order $k^2_y$, the $k_y$ dependence disappears for these values of $u$. 
Further, if  we expand Eq. (\ref{eq6_1}) in powers of $k_y$ we obtain
\begin{equation}
\begin{split}
	\ve &=  \sin(u/4)/(u/4)\, k_y - (2/u^5)\,\left[ u^3 \cos \right.(u/4)\\ &\left.+ 4 u^2 \sin(u/4) - 128 \sin^3(u/4)\right] k_y^3 + O(k_y^5),
\end{split}
\end{equation}
which is  linear in $k_y$, for small $k_y$, 
and the velocity becomes
\begin{equation}
	v_y/v_F= \partial \ve / \partial k_y \approx \sin(u/4)/(u/4).
\end{equation}
In Fig. \ref{fig6f} the velocities of the Dirac point, in the $x$ and $y$ directions, are given by the $j=0$ curves. 
For $u/2 = 2 j \pi$  we have
\begin{equation}\label{eq_approxfig6}
	\ve \approx \pm  
	k_y^3/8\,j^2\,\pi^2 + O(k_y^5),
\end{equation}
which is cubic in $k_y$  for small $k_y$. If  $j$ and consequently 
$u$ become larger  the dispersion gets flatter. In Fig. \ref{fig6}(b) we plot 
Eq. (\ref{eq_approxfig6}) for $u = 4 \pi$ and $u = 8 \pi$ as dashed curves, which correspond, respectively, to the zoomed-in plots of the 
solid red and dashed green curves of Fig. \ref{fig3}(b), shown here as blue and red curves.

\emph{Group velocity along the $y$-axis at the extra Dirac points.} 
The dispersion relation (\ref{eq6_1}) for the $k_y$ values of the extra Dirac 
points, determined by ${k_y}_{j,\pm} = \pm[u^2 / 4 - (2 j \pi)^2]^{1/2}$, gives 
us an idea of how dispersionless the spectrum near these points is along the 
$x$-direction. If ${k_y}_{j,\pm}$ exists, Eq. (\ref{eq6_1}) becomes
\begin{equation}
	\ve \approx 32\,\pi^2 j^2\sin(|k_x|/2)/u^2,
\end{equation}
and the partial derivative\cite{noimplicit2} of $\ve$ with respect to $k_x$ is
\begin{equation}\label{eq_vx}
	v_x /v_F= 
	 \partial \ve / \partial k_x \approx sign(k_x) 16\,\pi^2 j^2\cos(k_x/2)/u^2.
\end{equation}
This means that for smaller $j$ (the most distant extra Dirac points) the group 
velocity along the $x$-direction is strongly suppressed.
Further, as $u > 4 j \pi$ must hold  in order for $k_{y,j\pm}$ to be real, 
$|v_x|$ is 
smaller than $1\, (\equiv v_F) $ at $k_x = 0$. Only for the special values 
$u = 4 j \pi$,  for which new Dirac points appear, we have $|v_{jx}|=1$. 

Meanwhile the dispersion in the $k_y$ direction is also of interest. 
First, let us take $k_x = 0$ and expand the dispersion relation 
(\ref{eq6_1}) for $k_y - {k_y}_{j,\pm} \ll 1$. To first  order in this 
difference we obtain
\begin{equation}
	\ve \approx \pm [4 {k_y}_{j,\pm}^2 / u^2] (k_y - {k_y}_{j,\pm}).
\end{equation}
This gives  the velocity $v_y$ at the extra Dirac points
\begin{equation}\label{eq_vy}
	\frac{v_y}{v_F} = 
	\partial \ve / \partial k_y \approx 4 {k_y}^2_{j,\pm} / u^2 = 
	 4 \big[u^2 / 4 - 4 j^2 \pi^2 \big] / u^2.
\end{equation}
Since the coordinates of the extra Dirac points should be real, 
${k_y}^2_{j,\pm}$ is positive  and smaller than $u^2 / 4$ 
and we have $v_y < 1$ 
(the outer Dirac points, for $j = 1$, show the largest $v_y$). 
This entails that both $v_x$ and $v_y$ are renormalized  at the new Dirac 
points. 
A plot of the velocities of the extra Dirac points, in the $x$ and $y$ 
directions, given by Eqs (\ref{eq_vx}) and (\ref{eq_vy}), is shown in 
Fig. \ref{fig6f}. As seen, for the extra Dirac points, $v_{jx}$, shown by the 
dashed red curves, starts from $v_F$ and decreases to zero with 
increasing $u$ while $v_{jy}$ (dash-dotted blue curves), starts from zero and 
approaches $v_F$ slowly for large $u$. 
The thin solid black curves show the velocities for $j = 1$, obtained 
numerically from  Eq. (\ref{eq2a_6}). As can be seen, the two curves match almost perfectly.

\subsection{Density of states and conductivity}
\begin{figure}[ht]
  \begin{center}
	\includegraphics[height=4cm]{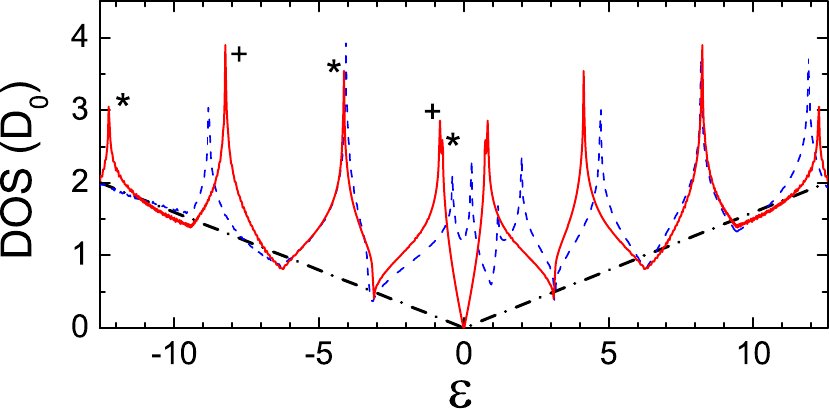} 
    \end{center}
	\vspace{-0.7cm}
  \caption{ (Color online) The DOS, for $u = 6 \pi$, and $W_b = 0.5$ and $0.4$ 
  is shown by, respectively, the solid red and dashed blue curve. 
  Stars and crosses placed near the peaks of the solid red curve 
  (for negative energies) are ascribed, respectively, to saddle points in the 
  spectrum, for 
  $k_x = 0$, and to minima for $k_x = \pi$. The DOS without a SL potential is 
  shown by the dash-dotted black curve.}\label{fig5}
\end{figure}

\emph{Density of states.} 
At zero temperature the density of states
(DOS) D(E) is given  by
\begin{equation}
	D(E)  = \sum_{n, {\bf k}} \delta(E - E_{n {\bf k}}),
\end{equation}
with $E$ the energy.  
We show the DOS in Fig. \ref{fig5}, for $W_b = 0.5$ 
(solid red curve) 
 and $W_b = 0.4$ 
(dashed blue curve), 
as well as the DOS for graphene without any SL potential 
(dash-dotted black curve); the latter is given by $D(\ve) = \ve D_0/ 2 \pi$, 
with $D_0 = L/\hbar v_F$ the amount of states per unit area and 
L is the period of the SL. The DOS shows an oscillating behaviour. 
The dips in it 
are located at the crossing 
points in the energy bands for $k_y = 0$ ($\ve = n \pi$), while the peaks marked 
by a star are ascribed to the saddle points between the crossing points for $k_x 
= 0$ and to the minima of the energy bands at the edge of the Brillouin zone, 
$k_x = \pm \pi$, marked by a cross. For $W_b = 0.4 \neq W_w$ the DOS 
(dashed blue curve) does not vanish at $\ve = 0$ nor is it symmetric about this 
energy. 

\emph{Conductivity tensor $\sigma$.} 
The diffusive dc conductivity $\sigma_{\mu \nu}$ for the SL system can be readily calculated from the spectrum if we assume a nearly constant relaxation time $\tau(E_F)\equiv \tau_F$. 
It is given by\cite{takis1}
\begin{equation}\label{eq_cond}
	\sigma_{\mu \nu}(E_F)=  
	\frac{e^2 \beta \tau_F}{A} 
	\sum_{n, \bf{k}}  
	v_{n\mu} v_{n\nu} f_{n\bf{k}}(1-f_{n\bf{k}}), 
\end{equation}
with $A$ the area of the system, $n$ the energy band index, $\mu,\,\nu = x,y,$ 
and $f_{n\bf{k}} = 1/[\exp(\beta(E_F - E_{n\bf{k}})) + 1]$ the 
equilibrium Fermi-Dirac distribution function; 
$E_F$ is the Fermi energy and $\beta = 1 /k_B T$.

\begin{figure}[ht]
  \begin{center}
  \subfigure{\includegraphics[width=4.2cm]{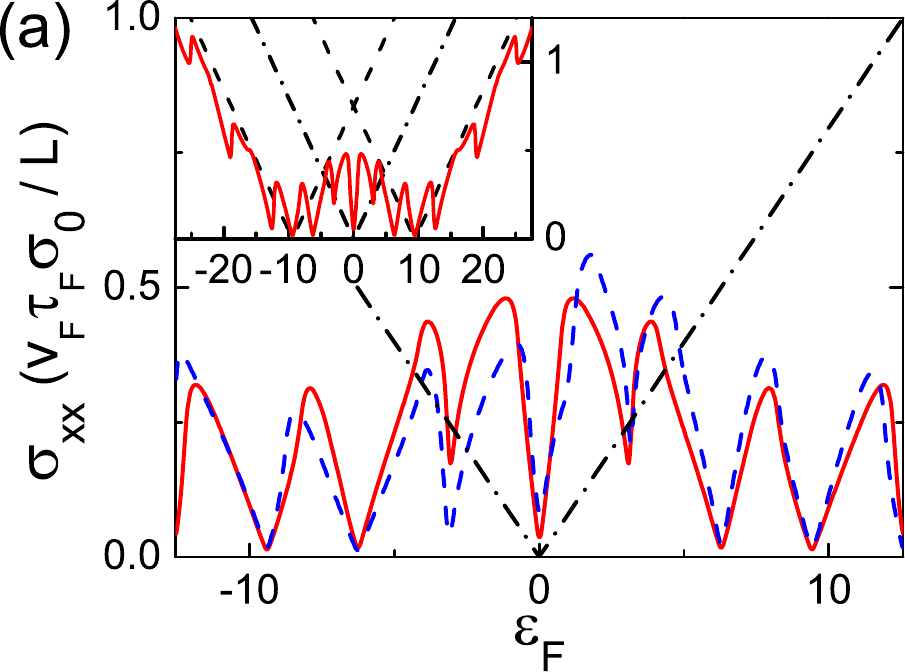}} 
  \subfigure{\includegraphics[width=4.2cm]{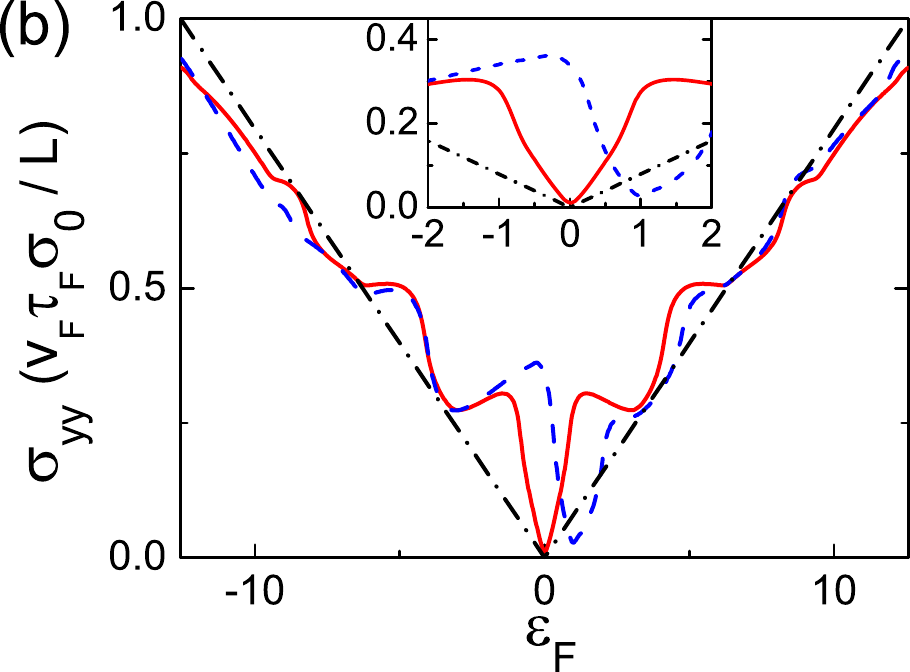}}\\ 
  \subfigure{\includegraphics[width=4.25cm]{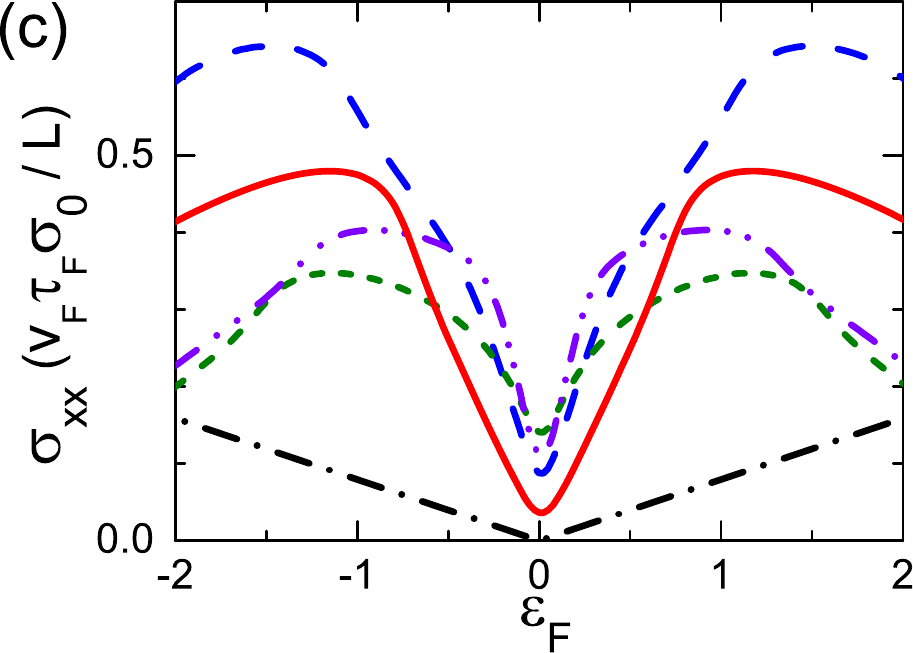}} 
  \subfigure{\includegraphics[width=4.25cm]{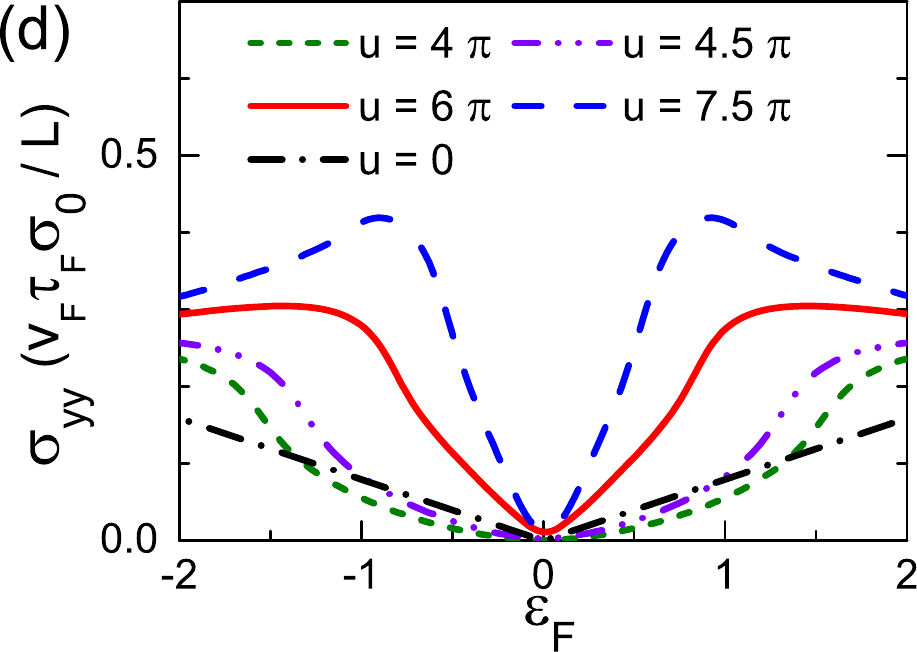}} 
  \end{center}
	\vspace{-0.7cm}
  \caption{(Color online) 
   (a) and (b) show the conductivities $\sigma_{xx}$ and $\sigma_{yy}$ vs Fermi 
   energy for a SL with $u = 6 \pi$. 
   The solid red and dashed blue curves are, respectively, for $W_b = 0.5$ and 
   $W_b = 0.4$ ($W_w = 1 - W_b$). The dash-dotted black curves show the 
   conductivities in the absence of the SL potential, 
   $\sigma_{xx} = \sigma_{yy} = \ve_F \sigma_0 / 4 \pi$. 
   The inset in (a) shows the 
   zoomed-out conductivity $\sigma_{xx}$, for $W_b = 0.5$, and 
   the dashed lines 
	are the conductivities of graphene in the absence of a SL but with a 
	constant non-zero potential applied, $-V_0/2$ and $+V_0/2$, so that 
 	$E_F$ is displaced by $V_0/2$ and $u/2 = 3 \pi$. 
   The inset in (b) is a zoom on $\sigma_{yy}$ for  small energies. 
   (c) and (d) show $\sigma_{xx}$ and $\sigma_{yy}$, for $W_b = W_w = 0.5$  for 
   different potential heights, such that $u = 4\pi,4.5\pi,6\pi,7.5\pi$, and 
   small energies.
  }\label{fig8}
\end{figure}

In Figs. \ref{fig8}(a) and  \ref{fig8}(b) we show,  respectively, $\sigma_{xx}$ 
and $\sigma_{yy}$ for a SL with $u = 6 \pi$, 
and the temperature dependence is given by $\beta = \hbar v_F / k_B T L = 20$ 
(in dimensionless units). 
The solid red and dashed blue curves correspond to $W_b = 0.5$ and $W_b = 0.4$, 
respectively. The dash-dotted black curve shows  
the conductivity at zero temperature and in the absence of a SL potential,  
$\sigma_{xx} = \sigma_{yy} = \ve_F \sigma_0 / 4 \pi,$ 
with $\ve_F= E_F L / \hbar v_F$ and 
 $\sigma_0 = e^2 / \hbar$.
Notice that $\sigma_{xx}$ is an oscillating function of the Fermi level  and 
recovers a quasi linear behaviour similar 
to that in graphene without a SL potential when 
$\ve_F$ is well above the barrier height, as shown in the inset, i.e. with the 
x axis displaced over the potential height or well depth, i.e., $u/2 = 3\pi$. 
On the average $\sigma_{yy}$  increases with $\ve_F$ and approaches the result 
without a SL for large energies. 

The oscillations in both $\sigma_{xx}$ and $\sigma_{yy}$ result from 
the motion of the Fermi level through the different SL minibands. Notice   that for $W_b = 0.4$ the conductivities are asymmetric with respect to electron and hole conduction. 
In both cases, $W_b = W_w = 0.5$ and $W_b = 0.4$, $\sigma_{xx}$ shows dips at 
$\ve_F = n\pi$, where energy band crossings occur in the spectrum for $k_y = 0$. 
In the former case the DOS has dips occurring at the same energy values 
that are 
dominated by the same crossings. In the latter  case we see that, unlike the 
DOS, $\sigma_{xx}$ is almost unaffected by the extra Dirac points for low 
energies since the spectrum is almost flat near these points. 
Similarly, for $W_b = 0.4$ we see that the minimum in $\sigma_{yy}$ is located 
at $\ve_F \approx 1$, 
that is,  the energy value for which  
the two extra Dirac points occur in the spectrum.

In Figs. \ref{fig8}(c) and  \ref{fig8}(d) we show, respectively, $\sigma_{xx}$ 
and $\sigma_{yy}$, for a  SL with $W_b = W_w = 0.5$, for different potential 
heights, such that $u = 0,4\pi,4.5\pi,6\pi,7.5\pi$, and 
$\beta = \hbar v_F / k_B T L = 20$. 
Notice that 
the conductivity $\sigma_{yy}$, in the low-energy range 
($\ve_F < 1$), is  lower than that 
in the absence of 
a SL potential while its slope increases as the potential barriers become 
higher. This is due to the extra Dirac points, that appear for larger 
potential heights, near which 
the velocity is larger 
along the $y$-axis. Notice that for $\ve_F < 1$ 
we have $\sigma_{xx} > \sigma_{yy}$ as a result of the inequality 
$v_x > v_y$ near the Dirac point.

\section{Conclusions}\label{sec4}
We investigated the appearance of zero modes, touching points at the Fermi level 
(extra Dirac points) in the spectrum of single-layer graphene in the presence of 
a 1D superlattice  
(SL). The system was described by a Dirac-type Hamiltonian, and the 
SL barriers were square.

In the general case of unequal well and barrier widths, there is no 
particle-hole symmetry and the extra Dirac points 
are no longer located at the Fermi level. 
We obtained  an analytical expression for the position 
of the crossing points in the spectrum. 
The extra ``$ $Dirac cones'' that appear at the various crossing points 
are reshaped, i.e., they are no longer circular symmetric and the slope is 
renormalized. For fixed height of the barriers, we found 
lower and upper bounds for the  barrier and well widths for the occurrence of these extra Dirac cones. 

For a  SL with  equal well and barrier widths 
we complemented the investigations of Refs. \onlinecite{parkwiggles} and 
\onlinecite{breywiggles}, which numerically demonstrated  the emergence of extra 
Dirac points (zero modes). In doing so we found a simple  analytical expression 
for the spatial distribution of these points in ${\bf k}$ space 
as well as a threshold value of the potential strength  for their appearance. 
Further, we approximated the dispersion relation for energies close to the Fermi 
energy  and found an explicit expression for the ${\bf k}$ space behaviour 
of the extra Dirac points at the Fermi level. Using this expression  
we showed how the group velocities at the various extra Dirac points are 
renormalized in the $x$ and $y$ directions. We also quantified how 
dispersionless the spectrum is in the neighbourhood of a Dirac point along the 
y-direction and the emergence of new 
points at which the conduction and the valence bands touch each other.

Finally, we obtained numerically the density of states (DOS), which exhibits 
an interesting oscillatory behaviour and 
is reflected in the conductivity of the system. We found 
that the dips in the DOS, for symmetric SLs, are located at the touching points 
in the spectrum for ${\bf k} = {\bf 0}$, i.e., for 
$\ve = n \pi$. For asymmetric SLs 
these dips persist but extra dips due to the extra Dirac points arise. The 
conductivity $\sigma_{xx}$ was found to have dips at the same values for $\ve_F$ 
as the DOS while 
the main features of $\sigma_{yy}$ in the low-energy range 
are due to the spectrum  near the extra Dirac points. 
We notice in passing that the influence of velocity renormalization
on transport was not studied in Refs. \onlinecite{parkwiggles} and \onlinecite{breywiggles} nor the modification
of the extra Dirac points for unequal well and barrier widths.

\begin{acknowledgments}
This work was supported by IMEC, the 
Flemish Science Foundation (FWO-Vl), the Belgian Science Policy 
(IAP), the Brazilian Council for Research (CNPq), and the Canadian 
NSERC Grant No. OGP0121756.
\end{acknowledgments}
\appendix
\section{Dispersion relation 
 for periodic systems}\label{app1}
The wave functions in the regions before and after the barrier, labeled, respectively, by $j = 1$ and $j= 2$, can be written as $\psi_j(x) = \Omega_j(x) \mathcal{A}_j$, with
\begin{equation}\label{eqapp1_1}
	\Omega = \matt{1}{1}{s e^{ i \phi}}{-s e^{- i \phi}}, \quad \mathcal{A} = \kvec{A}{B}
\end{equation}
Since the wave function of the entire  
periodic system is a Bloch function and the transfer matrix $\mathcal{T}$ connects the regions before and after the barriers, we have  
\begin{equation}\label{eqapp1_2}
	\psi(1) = e^{i k_x} \psi(0), \quad \mathcal{A}_1 = \mathcal{T} \mathcal{A}_2,
\end{equation}
with $k_x$ the Bloch wave vector. From these boundary conditions we extract the relation ($\lambda = [\ve^2 - k_y^2]^{1/2}$)
\begin{equation}\label{eqapp1_3}
	e^{-i k_x } e^{i \lambda \sigma_z} \mathcal{A}_2 = \mathcal{T} \mathcal{A}_2, \quad \text{ with } \mathcal{T} = \matt{w}{z}{z^*}{w^*}.
\end{equation}
For nontrivial solutions of Eq. (\ref{eqapp1_3}), the determinant of $\mathcal{A}_2 = \rvec{A,}{B}^T$ must be zero, i. e.,
\begin{equation}\label{eqapp1_4}
	\det\matt{e^{-i k_x} e^{i \lambda}-w}{-z}{-z^*}{e^{-i k_x} e^{-i \lambda} - w^*} = 0.  \end{equation}
Evaluating the determinant gives the dispersion relation
\begin{equation}\label{eqapp1_5}
	\cos k_x = \Re\{w e^{-i \lambda}\} =  
	\cos(\delta_t + \lambda)/|t|,
\end{equation}
with $1/w = t = |t| e^{i \delta_t}$.

\section{Crossing points for unequal barrier and well widths}\label{app2}
Suppose 
a solution $(\ve,k_x = 0,k_y)$ of the dispersion relation 
(\ref{eq2a_5}) is known for which the derivative $\partial \ve / \partial k_y$ 
at a certain $k_y$ value is undefined; then this 
$k_y$ value
can be a crossing point. The condition for such a solution is $\sin(\lambda W_w) = \sin(\Lambda W_b) = 0$, and $\cos(\lambda W_w) = \cos(\Lambda W_b) = \pm 1$, which entails  
\begin{equation}\label{eqapp2_1}
\begin{aligned}
	\lambda W_w &= j \pi,\\
	\Lambda W_b &= (j + 2 m) \pi,
\end{aligned}
\end{equation}
with $j$ and $m$ integers. Explicitly we obtain
\begin{equation}\label{eqapp2_2}
\begin{aligned}
	&\big[(\ve + u W_b)^2 - k_y^2\big]W_w^2 = (j \pi)^2,\\
	&\big[(\ve - u (1-W_b))^2 - k_y^2\big]W_b^2 = ((j + 2 m) \pi)^2.
\end{aligned}
\end{equation}
Subtracting the second equation from the first one in (B2)  gives  
\begin{equation}\label{eqapp2_3}
	 2 u \ve - u^2(1 - 2 W_b) =  \pi^2 \left(\frac{j^2}{W_w^2} - \frac{(j+2 m)^2}{W_b^2}\right),
\end{equation}
from which 
the corresponding value of the energy $\ve$ can be extracted. 
Substituting this value in  the first of Eqs. (\ref{eqapp2_2}) one obtains  
\begin{equation}\label{eqapp2_4}
\begin{aligned}
	& \ve_{j,m} = \frac{u}{2}(1-2 W_b) + \frac{\pi^2}{2 u}\left( \frac{j^2}{W_w^2}-\frac{(j+2 m)^2}{W_b^2}\right),\\
	& {k_y}_{j,m} = \pm  
	\big[(\ve_{j,m} + u W_b)^2 -(j \pi / W_w)^2\big]^{1/2}.
\end{aligned}
\end{equation}

\onecolumngrid
\vspace{1cm}
\twocolumngrid


\begin{thebibliography}{10}
\bibitem{hosummary}
In Ref. 9 a nearest-neighbour, tight-binding method was applied with  a large rectangular super-cell along a sinusoidal potential as basis. Extra touching points at the Fermi level were  observed and, together with the group velocity in their vicinity, their influence on  the density of states was demonstrated.
\bibitem{noimplicit2} The implicit function theorem cannot be applied if the gradient (jacobian) of the dispersion relation is zero. This occurs 
because of the crossing of the bands at these points:
from the form of the derivatives, too long to write down, we can see that they are both zero if $\sin(\lambda W_w) =\sin(\Lambda W_b) = 0$ while $\cos(\lambda W_w) = \cos(\Lambda W_b)$.
\end{thebibliography}
\end{document}